\documentclass[twocolumn,showpacs,prc,superscriptaddress]{revtex4}
\usepackage{amsmath}
\providecommand{\LyX}{L\kern-.1667em\lower.25em\hbox{Y}\kern-.125emX\@}

\usepackage[T1]{fontenc}
\usepackage[latin1]{inputenc}
\usepackage{graphics}
\makeatother
\begin{document}

\newcommand{\be}{\begin{eqnarray}}
\newcommand{\ee}{\end{eqnarray}}
\topmargin -1cm

\title{Multiple scattering effects in quasi free scattering from halo nuclei: 
a test to Distorted Wave Impulse Approximation}

\author{R. Crespo}
\affiliation{ Centro de F\'{\i}sica Nuclear, Universidade de Lisboa,
Av.\ Prof.\ Gama Pinto 2, 1649-003 Lisboa, Portugal}
\affiliation{ Departamento de F\'{\i}sica, Instituto Superior T\'ecnico, Taguspark,
Av. Prof. Cavaco  Silva,  Taguspark, 2780-990 Porto Salvo, Oeiras, Portugal}
\author{A. Deltuva}
\author{E. Cravo}
\author{M. Rodr\'{\i}guez-Gallardo}
\author{A.~C. Fonseca}
\affiliation{ Centro de F\'{\i}sica Nuclear, Universidade de Lisboa,
Av.\ Prof.\ Gama Pinto 2, 1649-003 Lisboa, Portugal}

%\email{raquel.crespo@tagus.ist.utl.pt//edgar@cii.fc.ul.pt
%//deltuva@cii.fc.ul.pt//mrodri@cii.fc.ul.pt//fonseca@cii.fc.ul.pt}

\date{\today}

\begin{abstract}
Full Faddeev-type calculations are performed for $^{11}$Be breakup on proton target
at 38.4, 100, and 200 MeV/u incident energies. The convergence of the multiple scattering
expansion is investigated. The results are compared with those of other 
frameworks like Distorted Wave Impulse Approximation that are
based on an incomplete and truncated multiple scattering expansion.
\end{abstract}

\maketitle
%24.50.+g 	Direct reactions 
%25.40.Ep 	Inelastic proton scattering

\section{Introduction}

Interest in quasi free breakup of a Halo nucleus stems mainly
from the structure information about the wave function of the struck valence
neutron.
Qualitatively, the quasi free scattering is a process
\cite{Jacob66,Jacob73} in which a particle knocks a nucleon
out of the nucleus (A) and no further strong interaction occurs between the
(A$_{n-1}$) nucleus and the two outgoing particles. 

Assuming that the nucleus (A) can be well described by an inert (A$_{n-1}$) core 
and a valence neutron, this reaction  can be studied using the few-body
Faddeev or equivalent Alt, Grassberger, and Sandhas 
(AGS) scattering framework \cite{faddeev60,Alt,Glockle} which can be viewed
as a multiple scattering series in terms of the transition amplitudes for each 
interacting pair. Both equations are consistent with the corresponding Schr\"odinger equation
and therefore provide an exact description of the quantum three-body problem
but are more suitable for the numerical solution.
The Faddeev equations are formulated for the components of the wave function,
while the AGS equations are integral equations for the transition operators
that lead more directly to the scattering observables.

 Traditionally  this reaction
 has been analysed using the Distorted Wave Impulse Approximation
(DWIA) scattering framework \cite{Chant77}. In this approach
one  assumes that the incoming particle
collides with the struck nucleon as if it was free, and therefore the cross
section for the quasi free scattering is given by the product of the
transition amplitude between the two colliding particles modulated by the wave
function of the nucleon inside the nucleus. Furthermore the relative motion
of the particles in the entrance and exit channels
is described by distorted waves. This approximation leads, in each order, to an incomplete
and truncated multiple scattering expansion.

In order to obtain accurate
information about the wave function, one needs to investigate the effect of
higher order rescattering between the two coliding particles and the 
(A$_{n-1}$) nucleus in the calculated knockout observables. In addition,
the contribution of the neglected terms in each order needs to be assessed.
This is the aim of the present work.

We consider here the quasi free scattering of the A nucleus
by a proton target leading to the breakup reaction p(A, A$_{n-1}$n)p. We take as an example the 
breakup of  $^{11}$Be halo into a proton target at 38.4, 100, and 200 MeV/u.

Recently the elastic scattering of $^{11}$Be from a proton target
was analysed using the Faddeev/AGS multiple scattering approach 
\cite{Crespo07b}. Its was shown that even at intermediate energies of 200 MeV/u
one needs to take into account 3rd order multiple scattering contributions
that are not included in the Glauber scattering framework
\cite{Crespo07a}.

In this work we briefly review the Faddeev/AGS multiple scattering framework
in section II. In section III we sumarize the DWIA formalism and identify
the terms of the corresponding multiple scattering expansion. In section IV we 
show the calculated fivefold breakup cross section using the Faddeev/AGS 
equations and compare with the results of the DWIA type.

\section{The Faddeev/AGS equations}

In this section, for completeness we briefly describe the Faddeev/AGS 
multiple scattering framework which treats all open channels in the
same footing.
Let us consider 3 particles (1,2,3) interacting by means of two-body
potentials. In this section we use 
the odd man out notation appropriate for 3-body problems
which means, for example, that the interaction between
the pair (2,3) is denoted as $v_1$. We assume that the system is 
non-relativistic and we write its total Hamiltonian as
\be
H = H_0 + \sum_\gamma v_\gamma \; ,
\ee
with the kinetic energy operator $H_0$ and
the interaction $v_\gamma$ for the pair $\gamma$.
The Hamiltonian can be rewritten as
\be
H = H_\alpha + V^\alpha \; ,
\ee
where $H_\alpha$ is the Hamiltonian for channel $\alpha$
\be
H_\alpha = H_0 + v_\alpha \; ,
\ee
and $V^\alpha$ represents the sum of interactions external to partition $\alpha$
\be
 V^\alpha = \sum_{\gamma \neq \alpha} v_\gamma \; .
\label{Valpha}
\ee
The $\alpha = 0$ partition corresponds to three free particles in the continuum where $V^0$ is the 
sum of all pair interactions. The total transition amplitude which describes the scattering from the 
initial state $\alpha$ to the final state $\beta$ is given in the post form 
(see Refs.~\cite{Glockle,Crespo07b} for a derivation)
\be
T_+^{\beta \alpha} = V^{\beta} + \sum_{\gamma \neq \alpha}
T_+^{\beta \gamma} G_0 t_{\gamma} \; 
\label{Tpost}
\ee
with the transition amplitude 
\be
 t_{\gamma} =  v_{\gamma} +  v_{\gamma} G_0  t_{\gamma} \; 
\ee 
and the free resolvent
\be
G_0 = (E+i0 - H_0)^{-1},
\ee
where $E$ is total energy of the three-particle system.
Equivalently in the prior form we write
\be
T_-^{\beta \alpha} = V^{\alpha} + \sum_{\gamma \neq \beta}
t_{\gamma}  G_0  T_-^{ \gamma   \alpha } \; .
\label{Tprior} 
\ee
One then defines the operators $U^{\beta \alpha}$ which are equivalent to
the transition amplitudes on the energy shell, such that
\be
U^{\beta \alpha} &=& \bar{\delta}_{\beta \alpha} G_\alpha^{-1}
+ T_+^{\beta \alpha}
\nonumber \\
&=& \bar{\delta}_{\beta \alpha} G_\beta^{-1}
+ T_-^{\beta \alpha} \; ,
\ee
where $ \bar{\delta}_{\beta \alpha} = 1 - {\delta}_{\beta \alpha}$.
From Eqs. (\ref{Tpost}-\ref{Tprior})
\be
U^{\beta \alpha} = \bar{\delta}_{\beta \alpha} G_0^{-1}
+ \sum_{\gamma} U^{\beta \gamma}G_0 t_\gamma \bar{\delta}_{\gamma \alpha} \; ,
\ee
or
\be
U^{\beta \alpha} = \bar{\delta}_{\beta \alpha} G_0^{-1}
+ \sum_{\gamma} \bar{\delta}_{\beta \gamma }
t_\gamma G_0 U^{\gamma \alpha} \; .
\label{Uba2}
\ee
These are the well known three-body AGS equations \cite{Alt} which for breakup 
($\beta = 0$ in the final state) become 
\be
U^{0 \alpha} = G^{-1}_0 + \sum_\gamma t_\gamma G_0 U^{\gamma \alpha}, 
\label{U0alpha}
\ee
where $U^{\gamma \alpha}$  is obtained from the solution of Eq.~(\ref{Uba2}) with 
$\alpha, \beta, \gamma = (1,2,3)$.  The scattering amplitudes are the matrix elements of 
$U^{\beta \alpha}$  calculated between initial and final states that are
eigenstates of the corresponding channel Hamiltonian $H_\alpha \, (H_\beta)$ with the same energy eigenvalue.
For elastic scattering and transfer those states are made up by the bound state wave function for pair 
$\alpha(\beta)$ times a relative plane wave between  particle $\alpha\, (\beta)$ and pair $\alpha \, (\beta)$. 
For breakup the final state is a product of two plane waves corresponding to the relative motion
of three free particles that may be expressed in any of the relative Jacobi variables. In the latter case the contribution of the  $G_0^{-1}$ term is zero.

\begin{figure}{\par\centering \resizebox*{0.1\textwidth}{!}
{\includegraphics{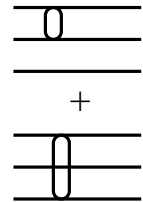}} \par}\caption{\label{Fig:single} 
Single scattering diagrams for breakup in the Faddeev scattering framework.}
\end{figure}

\begin{figure}{\par\centering \resizebox*{0.25\textwidth}{!}
{\includegraphics{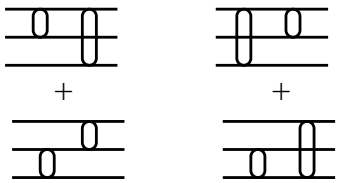}} \par}\caption{\label{Fig:double} 
Double scattering diagrams for breakup in the Faddeev scattering framework.}
\end{figure}

\begin{figure}{\par\centering \resizebox*{0.25\textwidth}{!}
{\includegraphics{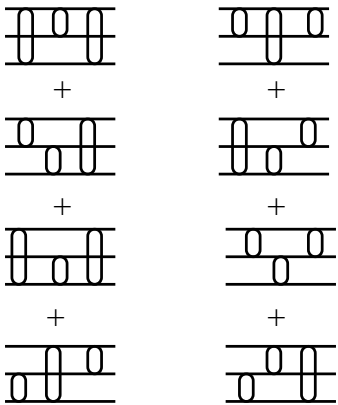}} \par}\caption{\label{Fig:triple} 
Triple scattering diagrams  for breakup in the Faddeev scattering framework.}
\end{figure}

The solution of the Faddeev/AGS equations can be found by iteration leading to
\be
U^{0 \alpha} &=& \sum_\gamma t_\gamma  
\bar{\delta}_{ \gamma  \alpha } + \sum_\gamma t_\gamma 
\sum_\xi G_0 \bar{\delta}_{\gamma \xi }  t_\xi
\bar{\delta}_{\xi  \alpha} \nonumber \\ &+&
\sum_\gamma t_\gamma 
\sum_\xi G_0  \bar{\delta}_{\gamma \xi }  t_\xi
\sum_\eta G_0 \bar{\delta}_{ \xi \eta} t_\eta  \bar{\delta}_{\eta \alpha}
 \nonumber \\ &+& \cdots \; ,
\label{AGS}
\ee
where the series is summed up by the Pad\'e method \cite{Pade}.
The successive terms of this series can be considered as first
order (single scattering), second order (double scattering) and so
on in the transition operators. The breakup series up to third order is represented diagramatically in Figs.~ \ref{Fig:single}
-  \ref{Fig:triple} where the upper particle is taken as particle 1 scattering from the 
bound state of the pair (23).

In our calculations Eqs.~(\ref{Uba2} - \ref{AGS}) are solved exactly after partial wave decomposition and discretization of all momentum variables. Since, in addition to the nuclear interaction between all three pairs, we include the Coulomb interaction between the proton and $^{10}$Be, we follow the technical developments implemented in Refs.~\cite{Del05b,Del06b} for proton-deuteron and $\alpha$-deuteron
elastic scattering and breakup that were also used in Ref.~\cite{Crespo07b} to study  p-$^{11}$Be elastic scattering. The treatment of the Coulomb interaction in the framework of three-body Faddeev/AGS equations is based on the method of screening of the Coulomb interaction plus renormalization~\cite{Del05b,Del06b}; it leads to fully converged results for the on-shell elastic, transfer and breakup observables that are independent of the choice of screening radius used in the calculations, as long as it is sufficiently large ($R \geq 10$ fm for the observables considered in the present work).

%%%%%%%%%%%%%%%%% single breakup
%%%%%%%%%%%%%%%%%%%%%%%%%%%%%%%%%%%%%%%%%%%%%%%%%%%%%%%%%%%%%%

%%%%%%%%%%%%%%%%% double breakup
%%%%%%%%%%%%%%%%%%%%%%%%%%%%%%%%%%%%%%%%%%%%%%%%%%%%%%%%%%%%%%

%%%%%%%%%%%%%%%%% triple breakup
%%%%%%%%%%%%%%%%%%%%%%%%%%%%%%%%%%%%%%%%%%%%%%%%%%%%%%%%%%%%%%

\section{The DWIA/PWIA}

\begin{figure}{\par\centering \resizebox*{0.1\textwidth}{!}
{\includegraphics{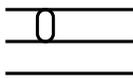}} \par}
\caption{\label{Fig:DWIA_single} 
Single scattering diagram in the DWIA scattering framework.}
\end{figure}

\begin{figure}{\par\centering \resizebox*{0.25\textwidth}{!}
{\includegraphics{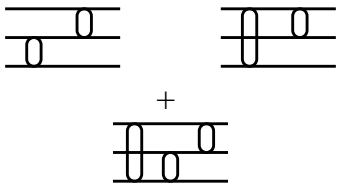}} \par}\caption
{\label{Fig:DWIA_exit} 
Diagrams due to the distortion in the exit channel 
in the DWIA scattering framework.}
\end{figure}

Let us consider the reaction $A(a,ab)B$ where an incident particle $a$ knocks
out a nucleon or a bound cluster $b$ in the target nucleus $A$ resulting in
three particles $(a,b,B)$ in the final state.
This reaction has been analyzed within the DWIA
as described for example in the work of Chant and Roos \cite{Chant77}. 
This is an approximate formalism and its validity in the application
to radioactive beam studies needs to be assessed.
In the work of reference \cite{Chant77} it is assumed that the projectile 
strucks the ejected particle freely and that the transition amplitude
is given by
\be
T_{AB} = \langle \eta^{(-)}_{Bab} | t_{ab} | \phi_{Bb} \eta^{(+)}_{aA} \rangle, 
\ee
where $\phi_{Bb}$ is the $(Bb)$ bound state wave function, 
and $\eta^{(+)}_{aA}$ and $\eta^{(-)}_{Bab}$  describe the relative
motion of the particles in the entrance and exit channels, respectively,
 neglecting the interaction $V_{ab}$. Therefore in 
 the entrance channel  $\eta^{(+)}_{aA}$ satisfies
\be
(T_{aA} + V_{aA} - V_{ab} - \epsilon_{aA}) \eta^{(+)}_{aA} = 0,
\ee
where $\epsilon_{aA}$ is the relative kinetic energy and $T_{aA}$ the relative
kinetic energy operator.
In standard DWIA calculations one takes $V_{aA}-V_{ab} \sim V_{aB}(\vec{r}_{aA})$.
For the exit channel one assumes as a good approximation to write
the 3-body wave function  $\eta^{(-)}_{Bab}$ as a factorized product
\be
\eta^{(-)}_{Bab} \sim \eta^{(-)}_{aB} \eta^{(-)}_{bB},
\ee
where
\be
(T_{aB} + V_{aB} -  \epsilon_{aB}) \eta^{(+)}_{aB} = 0,
\ee
and 
\be
(T_{bB} + V_{bB} -  \epsilon_{bB}) \eta^{(+)}_{bB} = 0.
\ee
The potentials $V_{aB}$ and $V_{bB}$ are taken to be the optical potentials
which describe the $a+B$ and $b+B$ scattering at energies 
$\epsilon_{aB}$ and $\epsilon_{bB}$ respectively.
Therefore one writes
\be
T_{AB} \sim \langle 
\eta^{(-)}_{aB} \eta^{(-)}_{bB} | t_{ab} | \phi_{Bb} \eta^{(+)}_{aA} \rangle.
\ee
If we now introduce plane waves in the exit channel
\be
T_{AB} \! \sim \! \langle 
\chi_{aB} \chi_{bB}
|(1 \!+ \!t_{aB} G_0)(1 \!+ \!t_{bB}G_0) t_{ab} | \phi_{Bb} \eta^{(+)}_{aA} \rangle\!,
\ee
we then obtain the truncated multiple scattering series
\be
t_{ab} +  t_{aB} G_0 t_{ab} + t_{bB} G_0 t_{ab}
+ t_{aB} G_0 t_{bB} G_0 t_{ab}.
\label{tdistortion}
\ee
The first order term is the single scattering
represented diagramatically in Fig.~\ref{Fig:DWIA_single}.
This is the only term retained in Plane Wave Impulse Approximation (PWIA)
calculations. The second and third order 
terms in Eq.~(\ref{tdistortion})  represented in 
Fig.~\ref{Fig:DWIA_exit} are due to distortion in the exit channel. 
The relative importance of these terms to the 
scattering can be assessed by performing Faddeev multiple scattering 
calculations. 

\begin{figure}{\par\centering \resizebox*{0.25\textwidth}{!}
{\includegraphics{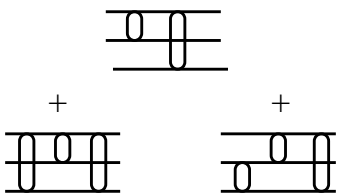}} \par}
\caption{\label{Fig:DWIA_income} 
Diagrams due to the distortion in the incoming channel 
in the DWIA scattering framework.}
\end{figure}

The distortion in the entrance channel is more difficult to bridge with
few-body calculations. In order to estimate this distortion in the case
of a light emited particle one could say for instance, that 
\be
\eta^{(+)}_{aA} \sim (1+G_0 t_{aB})\chi_{aA}.
\ee 
This leads to additional multiple scattering contributions
\be
t_{ab}G_0 t_{aB} +  t_{aB} G_0 t_{ab} G_0 t_{aB} +
 t_{bB} G_0 t_{ab}G_0 t_{aB} + {\cal O}(t^4).
\label{distortion-in}
\nonumber \\
\ee  
These terms are represented in Fig.~\ref{Fig:DWIA_income}.
We note that the distortion on the entrance and exit channels contribute
to a truncated and  incomplete multiple scattering series at each order. 

We concentrate on the study of the multiple scattering expansion.
In addition, we want to investigate if the single scattering term
becomes the dominant contribution as the energy increases.
 It was shown in the work of
Chant and Roos \cite{Chant77} that, contrary to what is expected,
the  DWIA results  do not converge to the PWIA results as the projectile 
energy increases. The importance of including the distortion in comparison with
the full Faddeev multiple scattering series is also
estimated in the present work. We refer
as DWIA-full a multiple scattering calculation that includes the multiple
scattering terms in Eqs.~(\ref{tdistortion}) and (\ref{distortion-in}). The calculations based on these equations follow exactly the same procedure as explained before for the solution of the Faddeev/AGS equations (\ref{Uba2}) through (\ref{AGS}), both in terms of initial and final states as well as treatment of Coulomb, so that one may obtain a consistent comparison between different terms in the multiple scattering representation of DWIA and Faddeev/AGS amplitudes.

We note that further approximations  are usually made in standard applications
of the DWIA when evaluating the transition amplitude, namely:
\begin{itemize}
\item
The potential approximation in the entrance channel
%of approximating 
$V_{aA}-V_{ab} \sim V_{aB}(\vec{r}_{aA})$.
\item
The factorization approximation, which is only exact in PWIA.
\item
On-shell approximation of the transition amplitude. 
\end{itemize}
We refer to this as DWIA-standard.
An attempt to estimate the effect of such approximations was done 
for example in the work of Ref. \cite{Kanayama} for (p,2p) reactions.
The Faddeev calculations and the DWIA-full 
do not make use of such approximations
in the summation of the multiple scattering expansion, and therefore
we are not addressing here the validity of such approximations which 
are responsible for additional shortcomings of DWIA-standard vis-a-vis the 
exact calculations.

%%%%%%%%%%%%%%%%% DWIA
%%%%%%%%%%%%%%%%%%%%%%%%%%%%%%%%%%%%%%%%%%%%%%%%%%%%%%%%%%%%%%

%%%%%%%%%%%%%%%%% DWIA

%%%%%%%%%%%%%%%%% DWIA
%%%%%%%%%%%%%%%%%%%%%%%%%%%%%%%%%%%%%%%%%%%%%%%%%%%%%%%%%%%%%%

\section{Results}

We now proceed to solve Faddeev/AGS equations (\ref{Uba2} - \ref{AGS}) and to
calculate the observables for the  proton-$^{11}$Be breakup.
As mentioned above the treatment of the Coulomb interaction,
as well as the calculational technique, is taken over from
Refs.~\cite{Del05b,Del06b}.

First we describe the pair  interactions p-n, p-$^{10}$Be and n-$^{10}$Be.
For p-n we take the realistic nucleon-nucleon CD Bonn potential \cite{CDBONN}.
The interaction between the valence neutron and the $^{10}$Be core is
assumed to be L-dependent as described in \cite{Crespo07b}.
For the potential between 
the proton and $^{10}$Be core we use a phemenological
optical model with parameters taken from the Watson global
optical potential parametrization \cite{Watson69,Crespo07b}. The energy dependent parameters 
of the optical potential are taken at the proton laboratory energy of the p-$^{11}$Be reaction
in inverse kinematics.

In the solution of the Faddeev equations we include n-p partial waves up to relative total angular momentum $\ell_{np} \leq 8$,
n-$^{10}$Be up to $\ell \leq 3$,
and p-$^{10}$Be up to   $L \leq 10 $.
Three-body total angular momentum is included up to 100.

%%%%%%%%%%%%%%%%%%%%%%%%%%%%%%%%%%%%%%%%%%%%
\begin{figure}
{\par\centering \resizebox*{0.45\textwidth}{!}
{\includegraphics{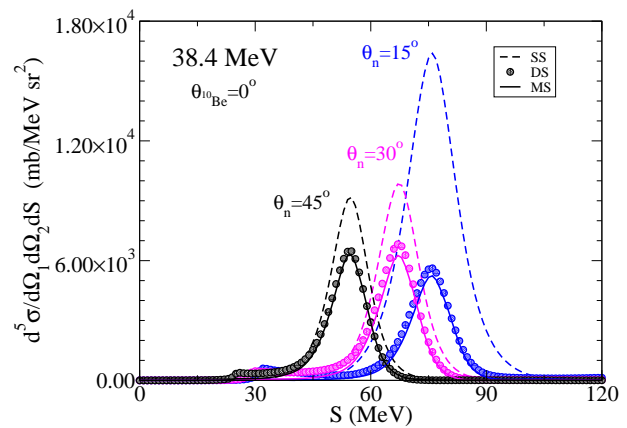}} \par}
\caption{\label{Fig:fivecross_msc_e38.4}
 (Color online) Cross section for the breakup
  $^{11}$Be(p,pn) at 38.4~MeV.
}
\end{figure}
%%%%%%%%%%%%%%%%%%%%%%%%%%%%%%%%%%%%%%%%%%%%

%%%%%%%%%%%%%%%%%%%%%%%%%%%%%%%%%%%%%%%%%%%%
\begin{figure}
{\par\centering \resizebox*{0.45\textwidth}{!}
{\includegraphics{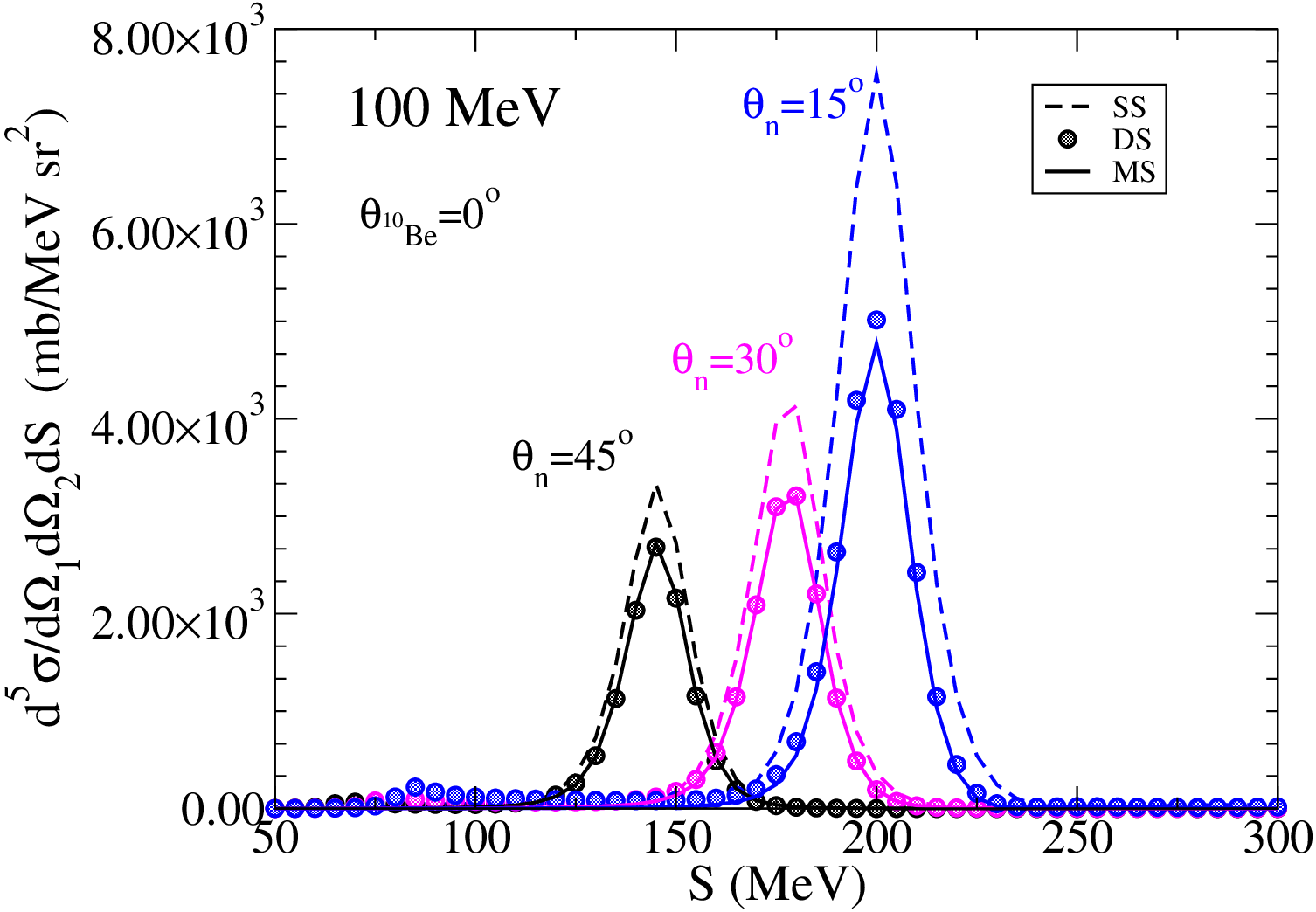}} \par}
\caption{\label{Fig:fivecross_msc_e100}
 (Color online) Cross section for the breakup
  $^{11}$Be(p,pn) at 100~MeV.
}
\end{figure}
%%%%%%%%%%%%%%%%%%%%%%%%%%%%%%%%%%%%%%%%%%%%

%%%%%%%%%%%%%%%%%%%%%%%%%%%%%%%%%%%%%%%%%%%%
\begin{figure}
{\par\centering \resizebox*{0.45\textwidth}{!}
{\includegraphics{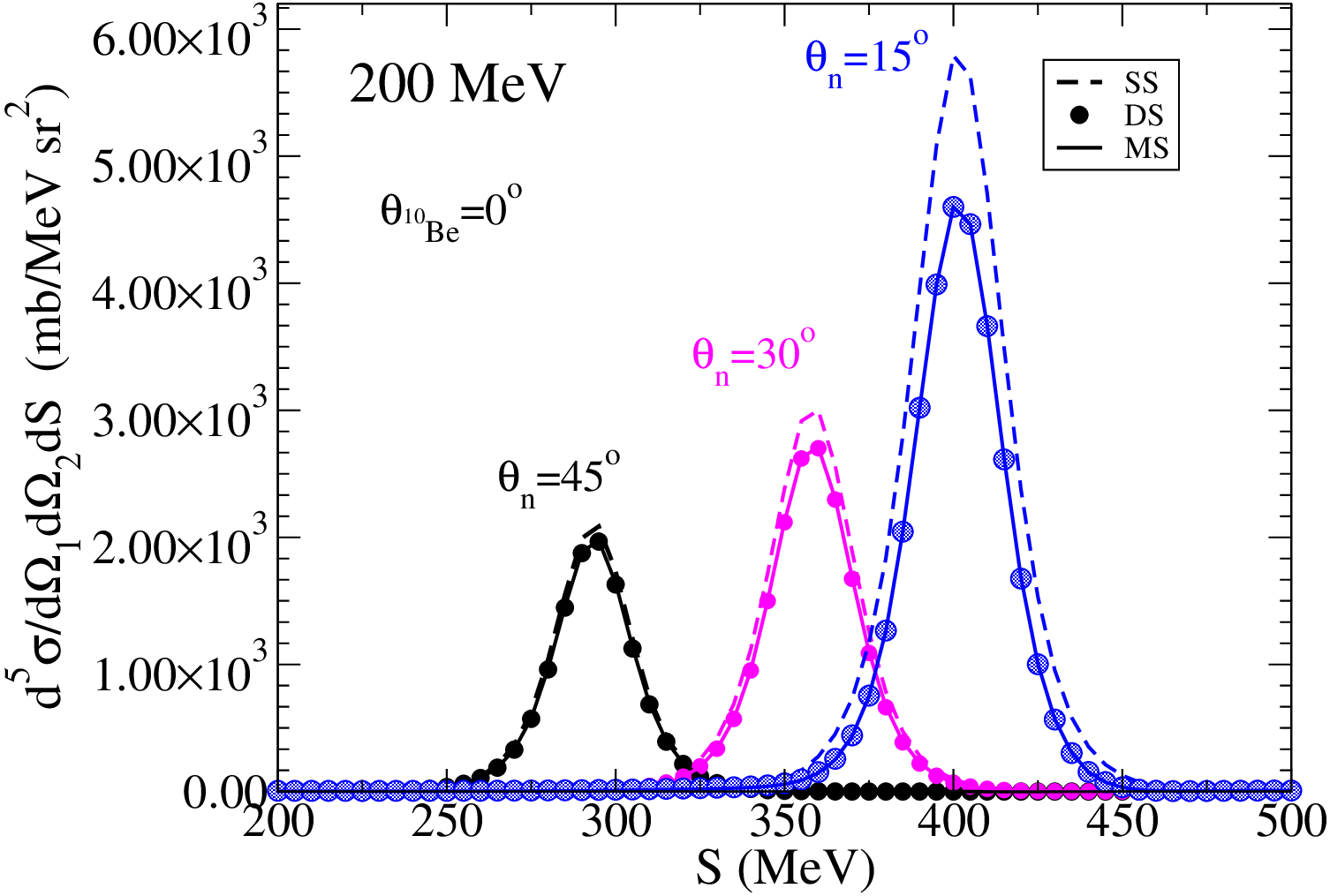}} \par}
\caption{\label{Fig:fivecross_msc_e200}
 (Color online) Cross section for the breakup
  $^{11}$Be(p,pn) at 200~MeV.
}
\end{figure}
%%%%%%%%%%%%%%%%%%%%%%%%%%%%%%%%%%%%%%%%%%%%

%%%%%%%%%%%%%%%%%%%%%%%%%%%%%%%%%%%%%%%%%%%%
\begin{figure}
{\par\centering \resizebox*{0.45\textwidth}{!}
{\includegraphics{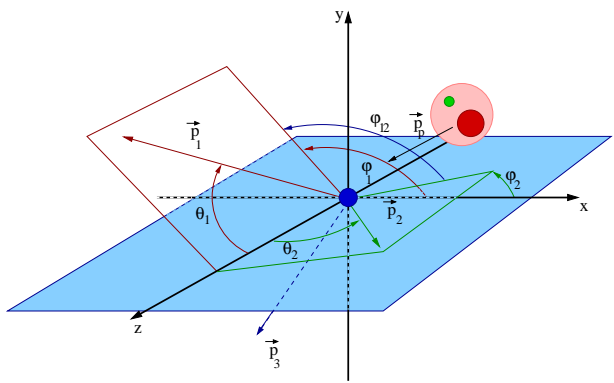}} \par}
\caption{\label{Fig:kinem}
 (Color online) Kinematic angles for breakup.
}
\end{figure}
%%%%%%%%%%%%%%%%%%%%%%%%%%%%%%%%%%%%%%%%%%%%

As a first result we show in 
Figs.~\ref{Fig:fivecross_msc_e38.4}--\ref{Fig:fivecross_msc_e200} 
the fivefold differential breakup cross section $ {d^5 \sigma}/{d\Omega_1 d\Omega_2 dS}$  
versus the arclength $S$ at $E_{\mathrm{lab}}/A = $ 38.4 MeV, 100 MeV, and 200 MeV respectively.
The kinematical configurations are characterized by the polar and azimuthal angles
$(\theta_i, \phi_i)$ of the two detected particles labeled 1 and 2 in  Fig.~\ref{Fig:kinem}.
We assume those particles to be the $^{10}$Be core (1) and the neutron (2).
We take three quasi free scattering configurations: (conf 1) with 
$(\theta_1, \phi_1)= (0^{\circ},0^{\circ})$ and  $(\theta_2, \phi_2)=(45^{\circ},180^{\circ})$; 
(conf 2) with $(\theta_1, \phi_1)= (0^{\circ},0^{\circ})$ and  
$(\theta_2, \phi_2)=(30^{\circ},180^{\circ})$; 
(conf 3) with $(\theta_1, \phi_1)= (0^{\circ},0^{\circ})$ 
and  $(\theta_2, \phi_2)=(15^{\circ},180^{\circ})$.
The arclength $S$ is related to the energies $E_i$ of the two detected particles 
in the standard way \cite{chmielewski:03a} as 
$S = \int_0^S dS$ with $dS = (dE_1^2 + dE_2^2)^{1/2}$.
In each configuration, the solid lines represent the full
multiple scattering (MS) results obtained from the solution of Faddeev/AGS equations.
The single scattering approximation (SS) represented by the dashed  lines 
is calculated taking into account the proton scattering from the struck nucleon
and the scattering from the valence core as represented by diagrams in 
Fig.~\ref{Fig:single}. As follows from  Fig.~\ref{Fig:fivecross_msc_e38.4}
the SS aproximation overestimates the exact result in any configuration at 38.4 MeV/u.
At 200 MeV/u, the SS
already provides a good representation of the full calculation in the case of
conf 1, but increasingly overestimates MS
 in conf 2 or conf 3 where the neutron is emitted at smaller angles.
Thus, contrary to what is predicted by Chant and Roos \cite{Chant77},
if the neutron is not detected at very forward angles, the single scattering
approaches the full calculation as the beam energy is increased.  
The double scattering (DS) approximation, represented by the circles, 
and calculated taking into account the diagrams of Figs.~\ref{Fig:single} and \ref{Fig:double},
reproduces well the full calculation at 200 MeV/u energy.
Therefore the Faddeev/AGS multiple scattering series for breakup has converged at second 
order level at this energy.

%%%%%%%%%%%%%%%%%%%%%%%%%%%%%%%%%%%%%%%%%%%%
\begin{figure}
{\par\centering \resizebox*{0.45\textwidth}{!}
{\includegraphics{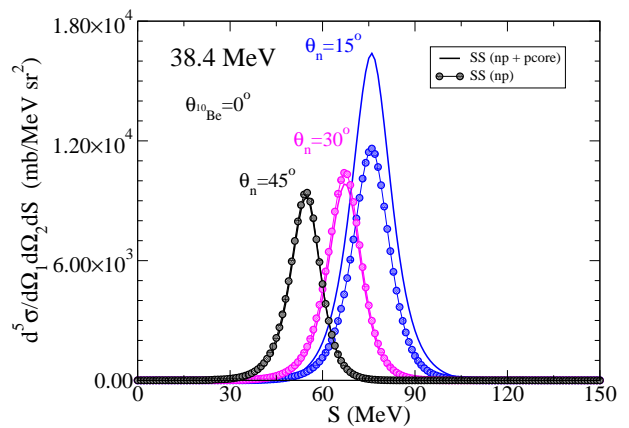}} \par}
\caption{\label{Fig:fivecross_ssca_38.4}
 (Color online) Cross section for the breakup
  $^{11}$Be(p,pn) at 38.4~MeV.
}
\end{figure}
%%%%%%%%%%%%%%%%%%%%%%%%%%%%%%%%%%%%%%%%%%%%

%%%%%%%%%%%%%%%%%%%%%%%%%%%%%%%%%%%%%%%%%%%%
\begin{figure}
{\par\centering \resizebox*{0.45\textwidth}{!}
{\includegraphics{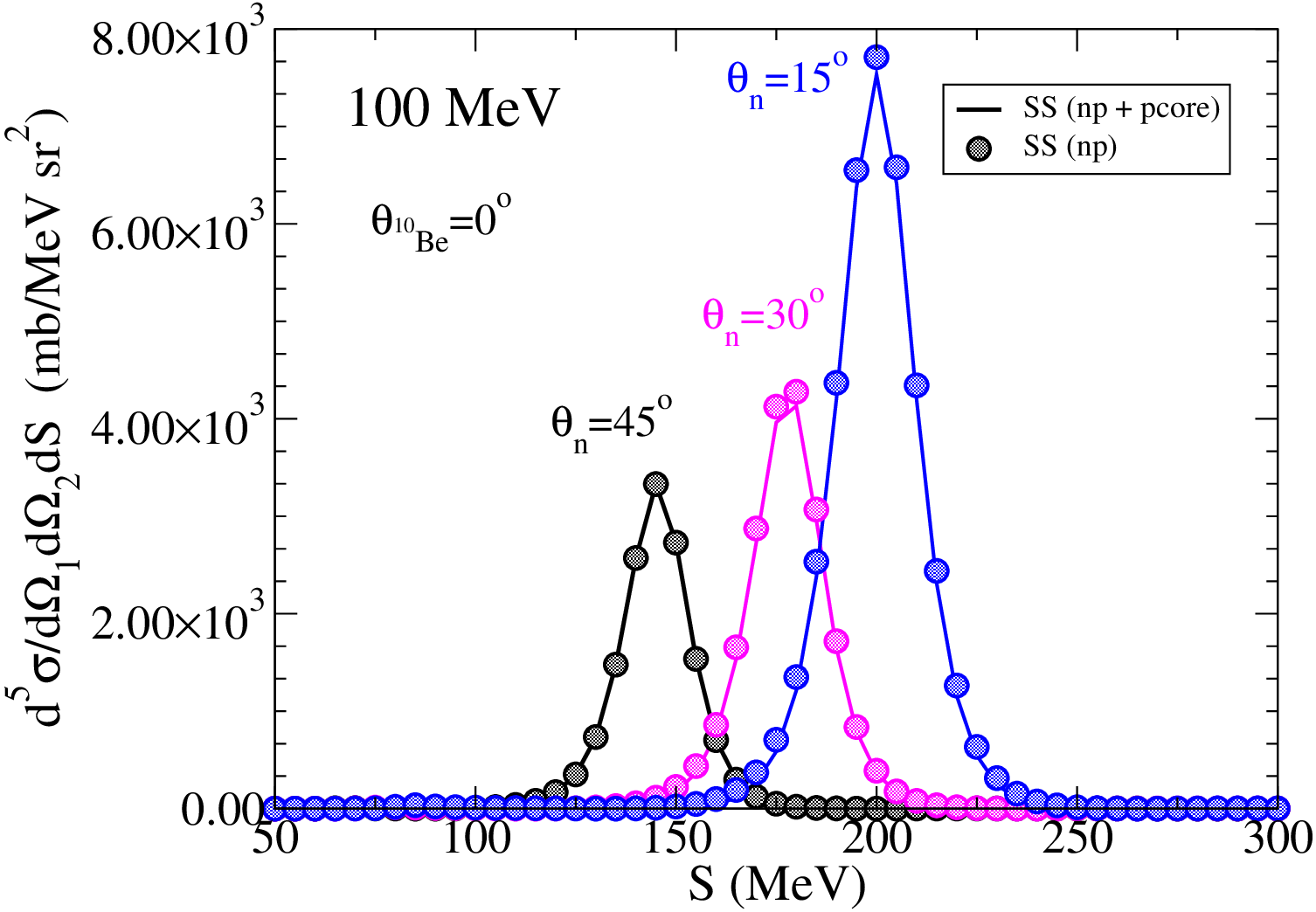}} \par}
\caption{\label{Fig:fivecross_ssca_e100}
 (Color online) Cross section for the breakup
  $^{11}$Be(p,pn) at 100~MeV.
}
\end{figure}
%%%%%%%%%%%%%%%%%%%%%%%%%%%%%%%%%%%%%%%%%%%%

%%%%%%%%%%%%%%%%%%%%%%%%%%%%%%%%%%%%%%%%%%%%
\begin{figure}
{\par\centering \resizebox*{0.45\textwidth}{!}
{\includegraphics{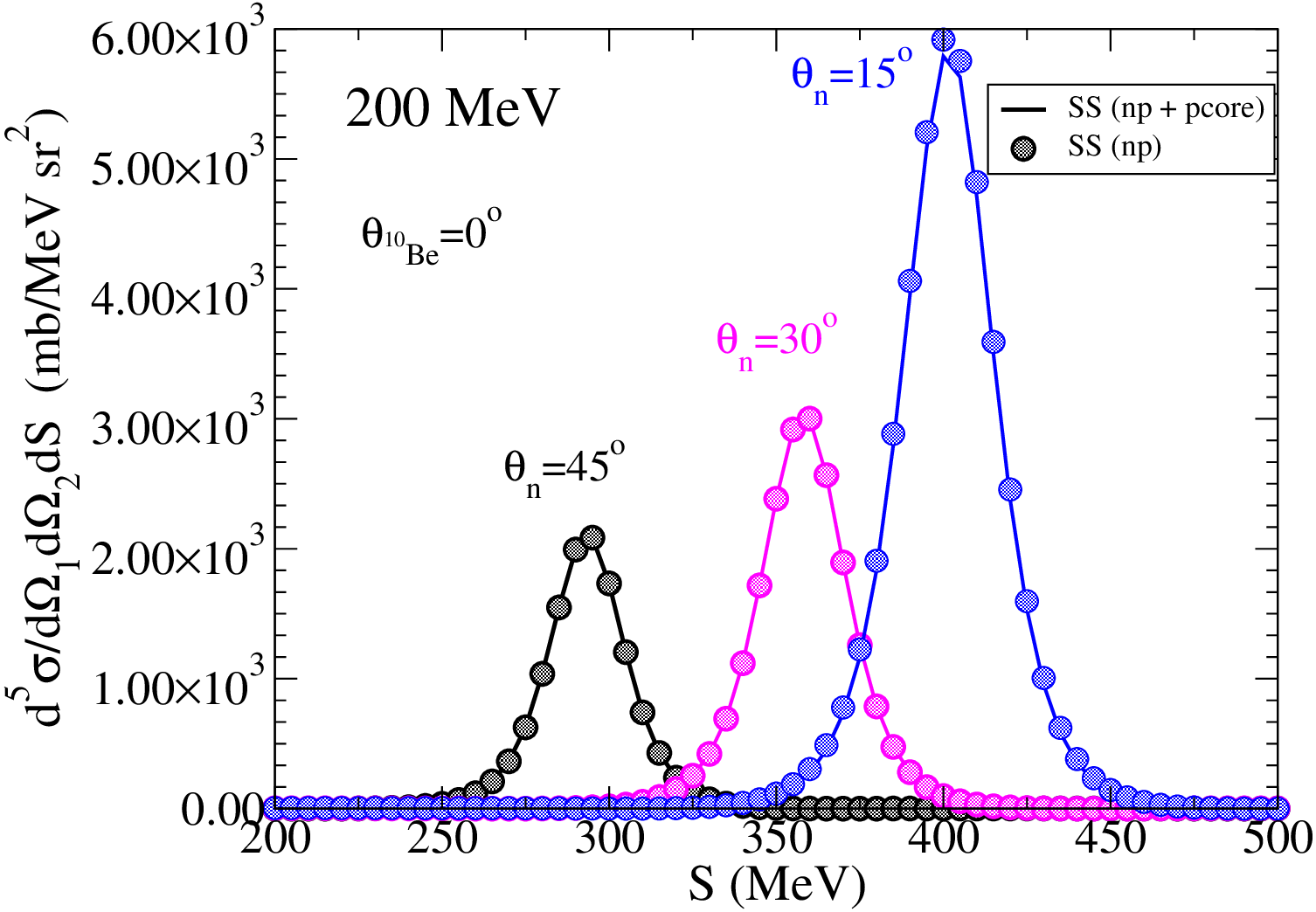}} \par}
\caption{\label{Fig:fivecross_ssca_e200}
 (Color online) Cross section for the breakup
  $^{11}$Be(p,pn) at 200~MeV.
}
\end{figure}
%%%%%%%%%%%%%%%%%%%%%%%%%%%%%%%%%%%%%%%%%%%%
Now we compare Faddeev/AGS and DWIA-full results for the breakup observables.
We begin by analyzing the single scattering term in
 Figs.~\ref{Fig:fivecross_ssca_38.4}--\ref{Fig:fivecross_ssca_e200}. 
The breakup cross section calculated taking into account both 
proton-neutron and proton-core contributions
is represented by the solid line. The circles correspond to the PWIA, i.e.,
without the proton-core scattering contribution.
At 38.4 MeV/u this contribution is very small
in the configurations 1 and 2, but is sizable 
when the neutron is emmited in the forward region. Therefore,
even at the single scattering level, calculations become
innacurate in some configurations at low energies. At high energies the
proton-core scattering contribution is negligible in all configurations.

\begin{figure}
{\par\centering \resizebox*{0.45\textwidth}{!}
{\includegraphics{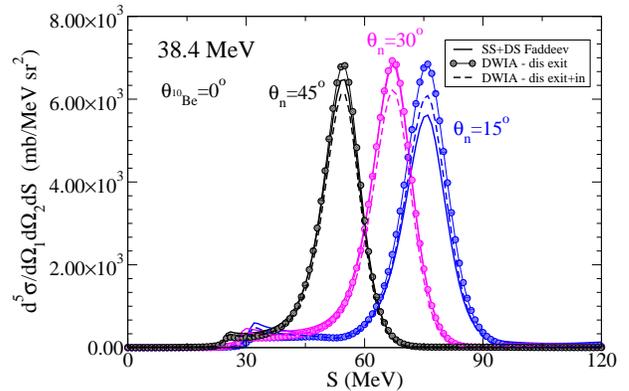}} \par}
\caption{\label{Fig:fivecross_dwia_e38.4}
 (Color online) Cross section for the breakup
  $^{11}$Be(p,pn) at 38.4~MeV.
}
\end{figure}

\begin{figure}
{\par\centering \resizebox*{0.45\textwidth}{!}
{\includegraphics{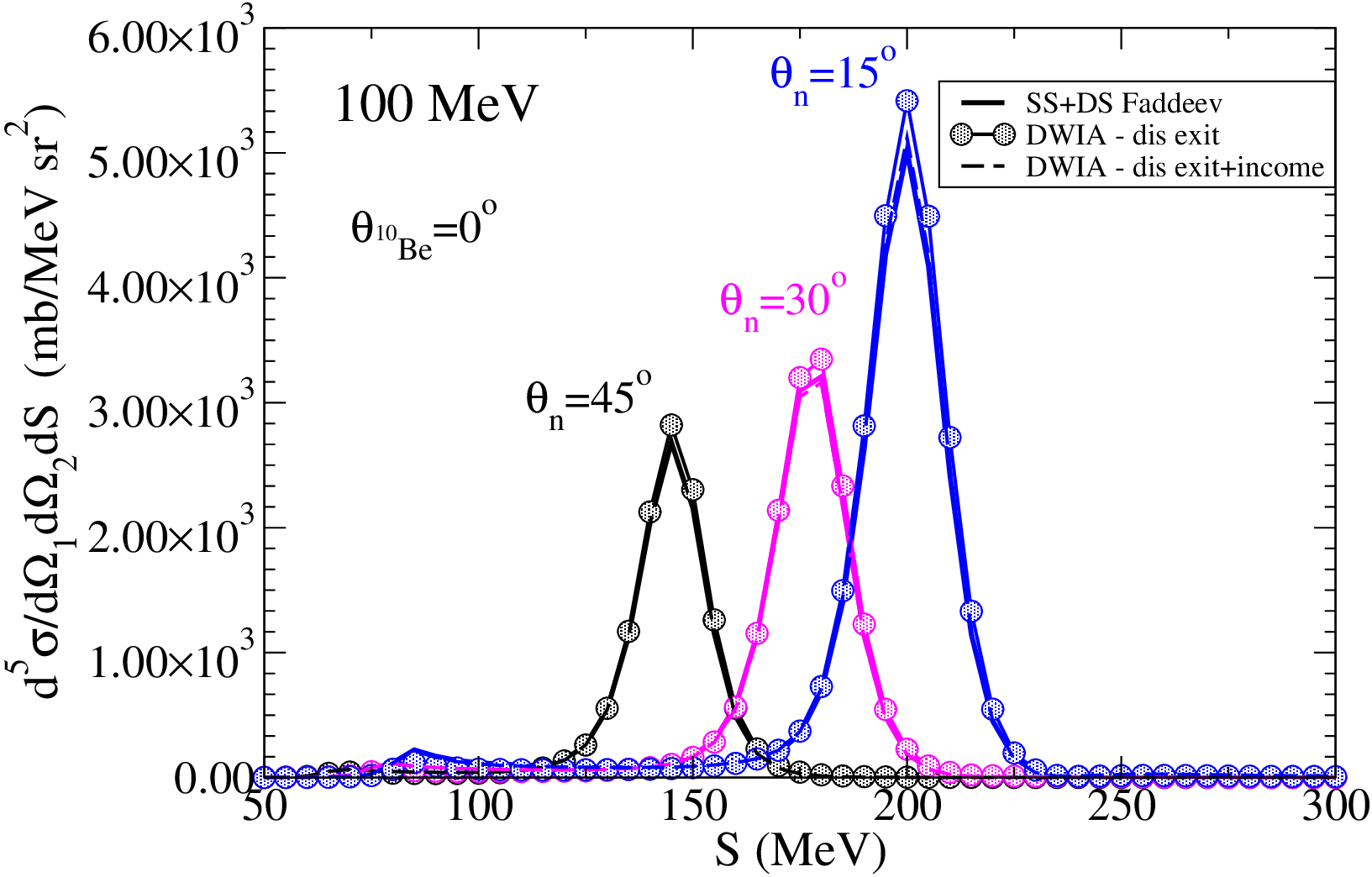}} \par}
\caption{\label{Fig:fivecross_dwia_e100}
 (Color online) Cross section for the breakup
  $^{11}$Be(p,pn) at 100~MeV.
}
\end{figure}

\begin{figure}
{\par\centering \resizebox*{0.45\textwidth}{!}
{\includegraphics{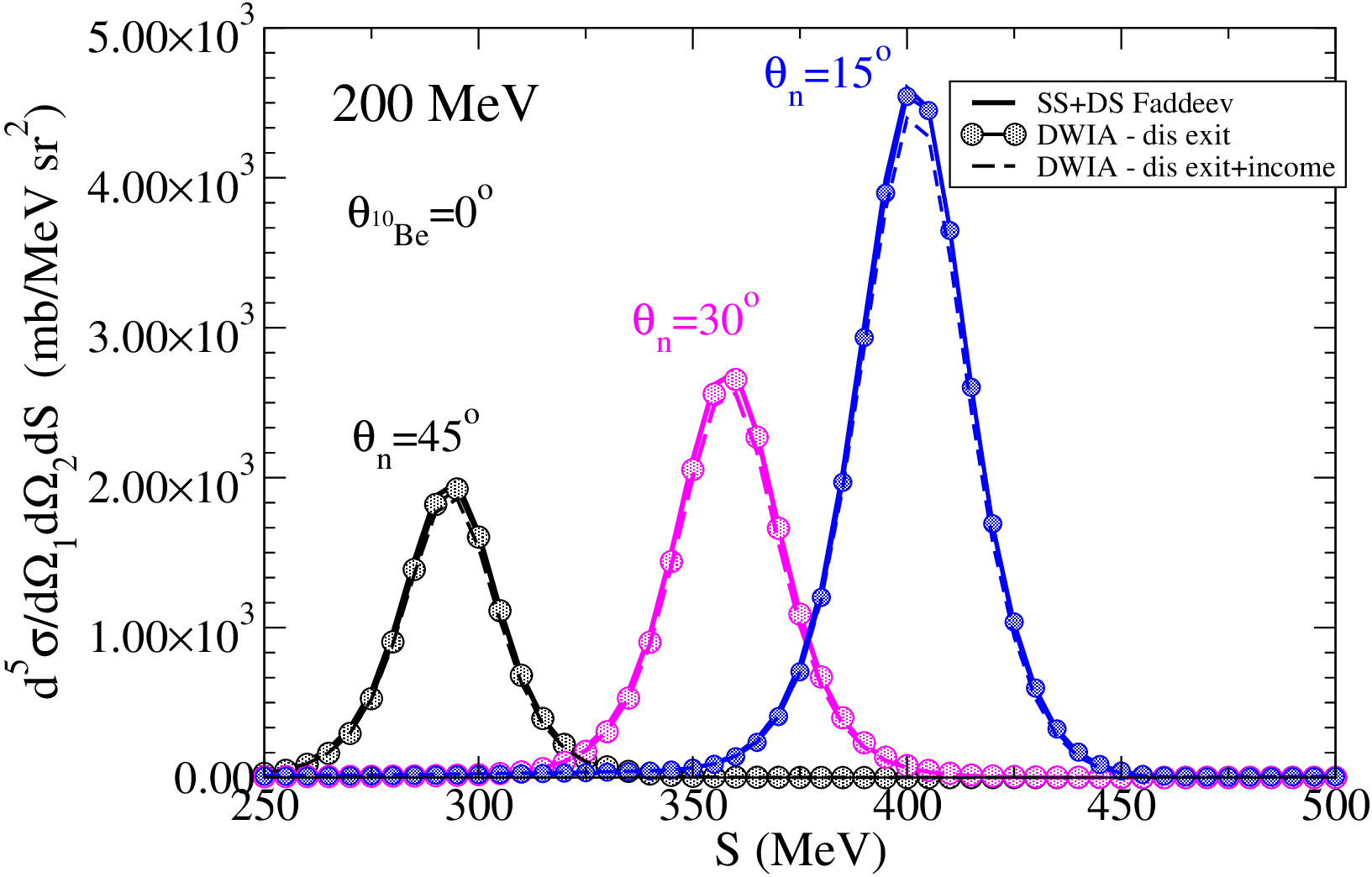}} \par}
\caption{\label{Fig:fivecross_dwia_e200}
 (Color online) Cross section for the breakup
  $^{11}$Be(p,pn) at 200~MeV.
}
\end{figure}

Since, as shown in Figs.~\ref{Fig:fivecross_msc_e38.4}--\ref{Fig:fivecross_msc_e200},
the DS results are very close to the full MS results, 
 we compare next in 
Figs.~\ref{Fig:fivecross_dwia_e38.4}--\ref{Fig:fivecross_dwia_e200}
the predictions of Faddeev/AGS and the DWIA-full approach including
in both calculations the corresponding
 multiple scattering terms up to second order. As explained before
the DWIA-full approach provides an incomplete series in all orders,
thought in first order the effects of the missing term are
negleagible at high energies as shown in 
Figs.~\ref{Fig:fivecross_ssca_38.4}--\ref{Fig:fivecross_ssca_e200}.

The Faddeev/AGS (solid line) results include all the single and double scattering
terms shown in Figs~\ref{Fig:single}--\ref{Fig:double}.
The DWIA-full assumes that the contribution of the scattering between
the proton and the neutron is dominant and therefore the single
scattering contribution due to the scattering of the proton from the
$^{10}$Be core is neglected. The circles include the first and second order
diagrams originated from the distortions in the exit channel
represented in Fig.~\ref{Fig:DWIA_exit}, and  the dashed 
curve includes, in addition, 
the second order diagrams originated from the distortion in
the incoming channel represented in 
Fig.~\ref{Fig:DWIA_income}. The DWIA-full results 
to second order (dashed line)
provide a poor approximation to the Faddeev/AGS  at 38.4 MeV
in any configuration. As the energy increases  DWIA-full approaches
the exact result. 
However one should keep in mind that DWIA-standard calculations assume 
further approximations as discussed in Sect.~III.

In Figs.~\ref{Fig:fivecross_msc_e200} 
and \ref{Fig:fivecross_ssca_38.4}--\ref{Fig:fivecross_ssca_e200}   
we show that in the high energy regime 
 PWIA approaches  the exact result
at least in some suitable breakup configurations.
Therefore, in this energy regime,
in order to get a reliable prediction of the breakup observables 
in quasi free scattering,  we
advocate that a better approach is to work with the PWIA scattering
framework within suitably chosen configurations
rather than in DWIA where uncontrolable approximations are made.
On the contrary, in PWIA the factorization is exact and the transition
amplitude for the scattering between the incident and knocked particle
is on the energy shell in the weak binding limit of the knocked
particle.

%%%%%%%%%%%%%%%%%%%%%%%%%%%%%%%%%%%%%%%%%%%%
%%%%%%%%%%%%%%%%%%%%%%%%%%%%%%%%%%%%%%%%%%%%

%%%%%%%%%%%%%%%%%%%%%%%%%%%%%%%%%%%%%%%%%%%%
%%%%%%%%%%%%%%%%%%%%%%%%%%%%%%%%%%%%%%%%%%%%

%%%%%%%%%%%%%%%%%%%%%%%%%%%%%%%%%%%%%%%%%%%%
%%%%%%%%%%%%%%%%%%%%%%%%%%%%%%%%%%%%%%%%%%%%

\section{Conclusions}

We have calculated the breakup cross section for $^{11}$Be breakup from
protons at 38.4, 100 and 200 MeV/u. Quasi free scattering conditions are considered in 
different configurations for the emitted neutron. The Faddeev/AGS formalism is used to show
that at these energies the multiple scattering expansion has converged at the second order.
We have also shown that at high incident energies
the single scattering (PWIA) approaches the full calculation in some
suitable configurations, namely when the neutron is not emitted at 
forward angles. We have also shown that the DWIA-full 
scattering framework provides  a poor approximation of the
Faddeev/AGS formalism at low energies, but approaches it at high energies.
However since standard applications make further use
of uncontrolable approximations such as the factorization and on-shell
approximations, one advocates that at these high energies 
it is adviseable to use the exact approach or
instead the PWIA in suitable configurations.

{\bf Acknowledgements:}
The work of R. Crespo is supported by the Funda\c c\~ao para a Ci\^encia e 
Tecnologia (FCT), through grant No.~POCTI/FNU/43421/2001. A. Deltuva is 
supported by the FCT grant SFRH/BPD/34628/2007 and all other authors by the
 FCT grant POCTI/ISFL/2/275.

\end{document}